**METHODOLOGY**    **Open Access**

# The whole is greater than the sum of its parts: improving music source separation by bridging networks

Ryosuke Sawata[1*] , Naoya Takahashi[1], Stefan Uhlich[2], Shusuke Takahashi[3] and Yuki Mitsufuji[1]

## Abstract

This paper presents the crossing scheme (X-scheme) for improving the performance of *deep neural network* (DNN)-based music source separation (MSS) with almost no increasing calculation cost. It consists of three components: (i) *multi-domain loss* (MDL), (ii) *bridging operation*, which couples the individual instrument networks, and (iii) *combination loss* (CL). MDL enables the taking advantage of the frequency- and time-domain representations of audio signals. We modify the target network, i.e., the network architecture of the original DNN-based MSS, by adding bridging paths for each output instrument to share their information. MDL is then applied to the combinations of the output sources as well as each independent source; hence, we called it CL. MDL and CL can easily be applied to many DNN-based separation methods as they are merely loss functions that are only used during training and do not affect the inference step. Bridging operation does not increase the number of learnable parameters in the network. Experimental results showed that the validity of Open-Unmix (UMX), densely connected dilated DenseNet (D3Net) and convolutional time-domain audio separation network (Conv-TasNet) extended with our X-scheme, respectively called X-UMX, X-D3Net and X-Conv-TasNet, by comparing them with their original versions. We also verified the effectiveness of X-scheme in a large-scale data regime, showing its generality with respect to data size. X-UMX Large (X-UMXL), which was trained on large-scale internal data and used in our experiments, is newly available at https://github.com/asteroid-team/asteroid/tree/master/egs/musdb18/X-UMX.

## 1 Introduction

There is a huge amount of music in our lives, e.g., from radio and TV, as background music in stores or provided by online streaming services [1–6]. To specialize music for diverse purposes, it is sometimes necessary to remix it, e.g., making the vocal tracks louder, suppressing undesired instruments, or upmixing to more audio channels. It is easy for us to implement such operations when we have access to each audio source independently that was used to mix the music. However, if we only have access to the final recording, which is often the case, this is much more challenging. In such cases, it is necessary to separate music into each instrument, which is called music source separation (MSS), to achieve the above operations.

MSS has a long history, and it is known to be a very challenging problem [7]; therefore, many approaches have been investigated, e.g., local Gaussian modeling [8, 9], non-negative matrix factorization [10–12], kernel additive modeling [13], and combinations of these approaches [14, 15]. Data-driven machine learning approaches for MSS have also been of great interest to researchers. Many methods that use *deep neural networks* (DNNs) have been investigated to improve MSS performance. Specifically, *multi-layer perceptrons* (MLPs) [16], *convolutional neural networks* (CNNs) [17], and *recurrent neural networks* (RNNs) [18], which are the three basic DNN architectures, have been used for MSS. An MLP was used to separate the input spectra then obtain separated results [19, 20]. CNNs and RNNs were used to achieve source separation with better quality [21–23] than previous MLP-based methods since the convolutional and recurrent layers of CNNs and RNNs can effectively capture the temporal contexts.

*Correspondence:
Ryosuke Sawata
Ryosuke.Sawata@sony.com
[1] Sony AI, Tokyo, Japan
[2] Sony Europe B.V., Stuttgart, Germany
[3] Sony Group Corporation, Tokyo, Japan

Although the above studies drastically improved MSS performance, there are two problems with respect to the training of music separation networks:

(P1) Most DNN-based MSS methods tend to handle only the time- or frequency-domain but not both.
(P2) They do not handle the mutual effect among output sources since network architectures and loss functions are independently computed for each estimated source and the corresponding ground truth.

For example, a well-known open-source MSS method, called *Open-Unmix* (UMX) [24][1], executes MSS only in the frequency-domain. It also applies the conventional *mean squared error* (MSE) loss function to individual pairs of estimated and corresponding ground truth magnitude spectrograms for each instrument. In other words, UMX trains networks individually for each instrument and achieves MSS by using all of each independent network one-by-one. In the field of speech enhancement (SE), which can be regarded as a case of audio source separation, there are methods for solving the above problems. For solving (P1), Kim et al. [25] showed the effectiveness of multi-domain processing via hybrid denoising networks, and Su et al. [26] reported that building two discriminators responsible for the time- and frequency-domains can enable effective denoising and dereverberation in their scheme of using *generative adversarial networks* (GANs). For solving (P2), from the classical SE methods such as Wiener filter [27] to current SE methods, e.g., noise-aware training [28] and noise-aware variational autoencoder [29], there are many situations in which knowing and using the information of the noise such as type, level, and time variation is generally beneficial for the following extraction of the target speech. In MSS, other non-target sources can be similarly regarded as "noise," and its information may be beneficial for the following target source separation. There is also research on using it for MSS using a Wiener filter [19], but it is used only as post-processing; thus, the information of other non-target sources is not used to train a DNN. Since our first work in [30], more models like Hybrid Demucs (HDmucs) and Hyblid Transformer Demucs (HTDemucs) [31, 32] as well as Band-split RNN (BSRNN) [33] appeared that showed the benefit of working jointly in both domains.

Inspired by these discussions, we first append an additional differentiable *short-time Fourier transform* (STFT) or *inverse STFT* (ISTFT) layer[2] during training only. To consider the characteristics on both of time and frequency domains, some existing methods such as [31] adopted the architecture having two separated branches each of which is respectively for time and frequency features, but it is the unique architecture per each method, and thus it is difficult to use its architecture for other method. On the other hand, the application of both loss functions in the time as well as frequency domains, i.e., applying *multi-domain loss* (MDL), easily becomes feasible for almost all existing methods since it is merely a loss function built on multi-domain. Intuitively, the two domains are also giving a complementary view of the separation performance. For a time-domain (TD)-loss, it might happen that we have a periodic noise pattern which is unnoticed by it as it is only computing an instantaneous error. However, in the frequency domain such a periodic noise pattern becomes visible and will be reduced by the frequency-domain (FD) loss. On the other hand, FD-loss lacks considering the effect caused by phase information since it only deals with magnitude spectrograms, but TD loss can contain it in the error of loss. Furthermore, to consider the relationship among output sources, we then bridge each instrument network by adding averaging operations if the original source separation is achieved by applying each independent instrument network to the input mixture. This is called bridging operation. For the bridged network to better determine the relationship among output instruments, we produce output spectrograms for instrument combinations and apply MDL to them. We call this loss computation combination loss (CL). The combination of bridging operation and CL helps the separation network determine the cause of an estimation error, i.e., which sources are leaking to the target instrument.

In summary, MDL solves (P1) since the separation network can help determine the estimation error in both time and frequency domains. Bridging operation and CL solve (P2) since they enable the separation network to handle the mutual relationships among the separated sources. We collectively call this "X-scheme," which crosses the information among all sources with MDL. It is important to note that X-scheme can improve the performance of DNN-based MSS systems while maintaining the original calculation cost. This is because MDL and CL only affect the training step and thus do not change the original inference step. Moreover, bridging operation requires only a slight network modification which does not increase the number of parameters that need to be learned and only slightly the computational costs. More specifically, the rate of computational cost that will be increased by applying X-scheme is depending on the original size of target network. However, as our X-scheme merely adds averaging operators to merge sub-networks together, these additional costs can often be neglected. For instance, only 4 additional averaging operators are needed in the case of a 4-instrument dataset like MUSDB18. No matter how small the deep neural networks are, we believe all existing ones should have much larger computational costs compared to adding a few averaging operators. Hence, there is almost no increase in computational cost by our proposed X-scheme.

---

[1] The implementations on two different libraries are available at https://github.com/sigsep/open-unmix-pytorch and https://github.com/sigsep/open-unmix-nnabla

[2] If the network outputs a spectrogram, we append an ISTFT layer whereas an STFT layer is added if the network output is a time signal.

Although we confirmed the validity of X-scheme in our previously proposed DNN-based MSS method, i.e., extended UMX (X-UMX) [30] realized by applying X-scheme to UMX, there remains three questions: (i) its generality to other types of network architectures, (ii) the effective positions where we should bridge the paths of the target networks, and (iii) its scalability to a large-scale data regime. Hence, in this paper, we address these questions. Specifically, we validate the effectiveness of X-scheme by applying it to different types of DNN-based MSS methods: well-known CNN-based and RNN-based ones, i.e., *densely connected dilated DenseNet* (D3Net) [34, 35] and *Open-Unmix* (UMX) [24]. Furthermore, not only these frequency-domain networks (i.e., UMX and D3Net) but also well-known time-domain one, convolutional time-domain audio separation network (Conv-TasNet) [1], is extended by X-scheme in this paper. We also present a detailed study regarding the bridging positions and potential to use a large dataset for training X-UMX.

The rest of this paper is organized as follows. In Section 2, we give a brief review of related work. In Section 3, we present X-scheme. In Section 4, we show the effectiveness of X-scheme by applying it to UMX, D3Net, and Conv-TasNet resulting in X-UMX, X-D3Net, and X-Conv-TasNet in terms of the MSS task. Finally, we conclude this paper in Section 5.

## 2 Related work

DNN-based MSS methods can be roughly categorized into time- and frequency-domain methods. UMX [24] receives the input spectrogram of a mixture song and extracts the target instrument by using fully connected and bi-directional long short-term memory (BLSTM) layers on the spectrogram, i.e., it works in the frequency domain. Similarly, D3Net [34, 35] extracts the target instrument from the input spectrogram by using convolutional layers in the frequency-domain. Note that they use a *multichannel Wiener filter* (MWF) [19] to reduce artifacts caused by non-linear separation due to DNN-based processing. Frequency-domain-based methods are powerful; thus, both methods recorded good scores on MUSDB18, which is a public dataset prepared for signal separation evaluation campaign (SiSEC) 2018 [36].

Time-domain methods directly operate on time-domain signals. To the best of our knowledge, Lluís et al. and Stoller et al. almost simultaneously started to explore time-domain MSS methods [37, 38]. However, the MSS performances of such methods were inferior to those of frequency-domain based methods. Specifically, the overall *signal-to-distortion ratio* (SDR) was reported to be only around 3.2 dB, which was almost 2 dB behind that of frequency-domain methods. Note that the experiments in the above studies were conducted on the same public dataset, i.e., MUSDB18; thus, we can compare their results. Défossez et al. then investigated a new time-domain method, Demucs [39], which is based on Wave-U-net [38]. Demucus improves the modeling capability by incorporating gated linear unit layers [40], BLSTM, and faster strided convolutions; thus, it demonstrated competitive results to frequency-domain methods on MUSDB18.

Although both time- and frequency-domain methods have recorded good MSS performance, there are still concerns. Almost all DNN-based frequency-domain methods tend to use only a spectrogram without phase information since it is difficult for DNNs to work with complex data. The phase information is often ignored with such methods. Therefore, the phase of the input mixture is often used with the output magnitude spectrogram to be able to compute the ISTFT, although this might yield a mismatch to the target source's spectrogram. The Fourier basis, which is used to calculate the above spectrogram, is not always optimal for DNN-based MSS methods. Time-domain methods, however, can optimize their networks from the perspective of being end-to-end, i.e., including the phase information, but tend to make the training more difficult. Inspired by this insight, we previously proposed X-UMX, which can use time-domain information via MDL [30], and confirmed that it performed better than UMX. Methods using time and frequency information in a hybrid manner were proposed for DNN-based MSS. For example, KUIELAB-MDX-Net [41] and Danna-Sep [42] are hybrid methods using time and frequency features. Specifically, they combine the heterogeneous time- and frequency-based MSS networks on the basis of the blending scheme [22], resulting in high performing hybrid MSS.

The number of methods using complex-valued features, i.e., spectrogram magnitude and the corresponding phase via STFT, for MSS has recently been increasing [43–45]. Specifically, *latent source attentive frequency transformation* (LaSAFT) [43] and its light version, LightSAFT [44], use *complex-as-channels* (CaC) [46] built on U-net [47], enabling MSS in the complex-valued domain. Défossez et al. also improved upon the original Demucs by using CaC, called HDemucs [31], to use time as well as complex-valued frequency information. Its architecture consists of two branches were each handles either time or complex-valued frequency input, respectively. Liu et al. proposed *channel-wise subband phase-aware ResUNet* (CWS-PResUNet) [45] which includes phase estimation by using the loss function of *complex ideal ratio mask* (cIRM) [48]. Their motivations, which involve phase information as well as spectrogram magnitude, are similar to those of the hybrid methods that compensate for the missing phase information by adding the time-domain signal. Therefore, the above complex-domain methods are hybrid methods.

There have been several attempts to directly estimate the phase of the target source [49, 50]. PhaseNet [49] successfully predicts the phase information by defining the phase-estimation problem as a classification of discretized phase values. DiffPhase [51] generates as well as predicts the phase through the framework of a diffusion-based generative model, which is suitable for the given spectrogram magnitude. The authors reported that the perceptual scores of reconstructed time signals were high even when their phases were partially generated.

From this literature review, we can see that using the time domain or similar features as well as the frequency domain is important to achieve good MSS performance. However, changing the network architecture such that the time- and frequency-domain features input can be jointly used and optimizing this new architecture may be a laborious task. X-scheme, which includes MDL, is simple and easy to use, thus it enables many methods to handle both time- as well as frequency-domain features in a hybrid manner.

Furthermore, there are some studies that attempted to integrate sub-networks each of which is dedicated for extracting one specific instrument [3–5]. Specifically, Meseguer-Brocal et al. proposed DNN-based MSS method that used just a single conditioned network [3].

By applying Feature-wise Linear Modulation (FiLM) [6] to the target network as conditioning, separating an arbitrary desired instrument through a single network becomes feasible. Selecting which instrument should be separated is achieved by FiLM-based conditioning without instrument-wise training. Furthermore, Slizovskaia et al. proposed the conditioned network-based MSS method that accepts visual features as well as audio ones, i.e., audio-visual features [4]. Besides using the conditioned network, Kadandale et al. proposed multi-task model-based MSS method that used the unified single network outputting all instruments simultaneously [5]. While using a conditioning scheme might require a longer network training to ensure that we do enough training steps for each conditioning signal, multi-task model-based networks need fewer iterations due to simultaneously learning them as multi-task. They increase the number of output instruments by changing the number of output kernels of U-net and then they easily change the each instrument's dedicated network to the unified multi-task one. However, their method only tried on CNN-based method, i.e., U-net, and it might be difficult to apply to other types of DNNs.

Our X-scheme can be regarded as a modification to change the target network to a multi-task one by bridging, and it can further be applied to not only CNN-based but also other types of networks as shown in the following sections.

## 3 X-scheme for DNN-based MSS

In this section, we describe X-scheme, which consists of three components, i.e., MDL, bridging operation, and CL. As mentioned in Section 1, MDL should solve (P1) and bridging operation and CL should solve (P2).

Throughout the paper, we use the following notations. We first assume that the time-domain mixture signal $x$ consists of $J$ sources, i.e.,

$$x = \sum_{j=1}^{J} y_j, \qquad (1)$$

where $y_j$ denotes the time-domain signal of the $j$th source. Note that $x$ and $y_j$ are column vectors with their samples, which they respectively denote the monaural signals. In general, the audio signal of music consists of two channels, i.e., stereo signal. However, the calculation of some metrics such as MSE and SDR that are used in our method does not have unique operators specialized for the multi-channel signal, and thus we calculated the following loss values by using each channel one-by-one and summed up them resulting in the final loss. To the best of our knowledge, although there is a multichannel version of classical SDR (e.g., https://github.com/sigsep/

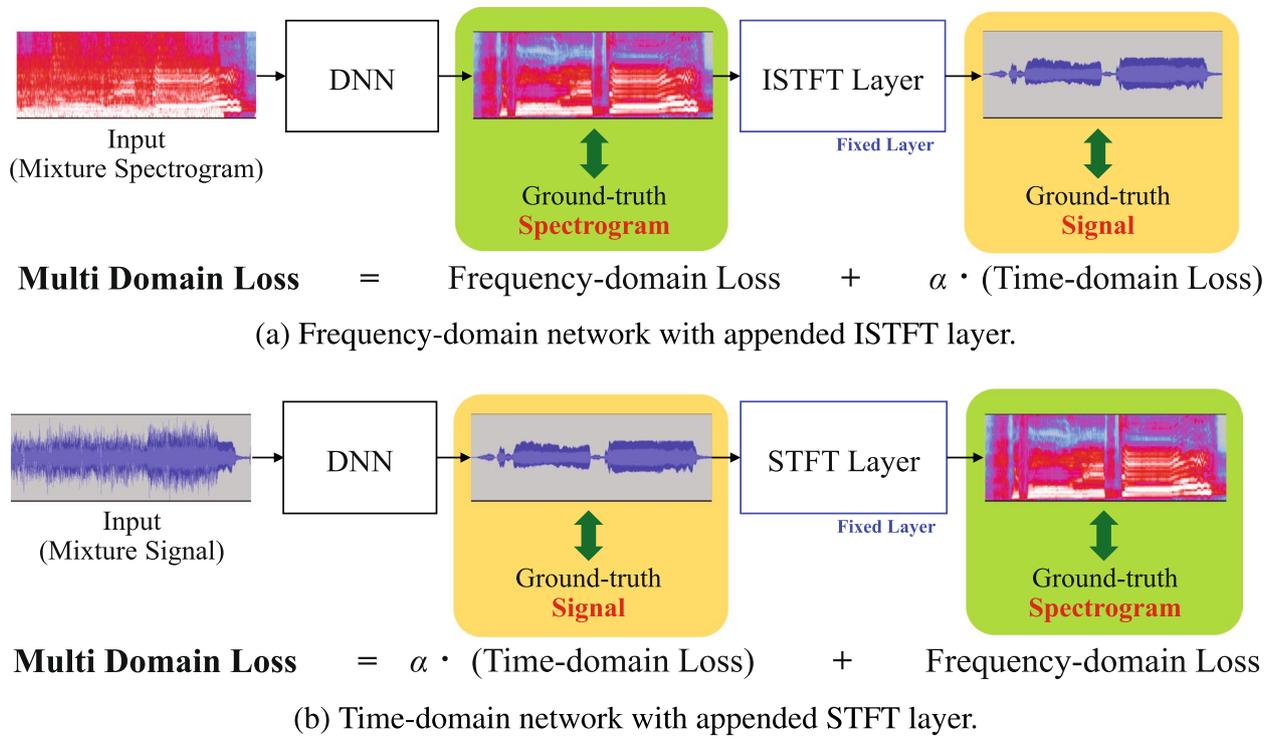

**Fig. 1** Multi -domain loss (MDL). Note that $\alpha$ is scaling parameter

bsseval), almost all of other existing methods and their implementation also handled each channel of stereo signals one-by-one. Thus, for sake of simplicity, the vectors used in the following equations are denotes as single, i.e., monaural, signals. The DNN then predicts the spectrogram of the $j$th target source from the input mixture spectrogram $X = \mathcal{S}\{x\}$:

$$\hat{y}_j = \mathcal{S}^{-1}\{\hat{Y}_j\}, \qquad (2)$$

where $\mathcal{S}$ and $\mathcal{S}^{-1}$ represent the operators of STFT and inverse STFT (ISTFT), respectively. A variable with the hat symbol, e.g., $\hat{\bullet}$, denotes the results estimated with the DNN. Therefore, $\hat{y}_j$ and $\hat{Y}_j$ are respectively the predicted time- and frequency-domain results of ground truths, i.e., $y_j$ and $Y_j$, via the DNN.

### 3.1 Multi-domain loss

For MDL, we first append an additional differentiable and fixed STFT or ISTFT layer after the final layer of the target DNN, as shown in Fig. 1. STFT and ISTFT consist of only product-sum operation, called butterfly computation [2], and thus all computational operations of it are differentiable. In other words, STFT and ISTFT consist of just some matrix-vector products each of which is differentiable. It is then possible to calculate the loss functions in both time- and frequency-domains before and after the appended layer. Hence, we can easily add STFT and ISTFT as the differentiable operators resulting in STFT and ISTFT layers. Since this appended layer is only used during training for computing MDL, it does not affect the inference step. In X-scheme, we use the loss functions of the MSE and *weighted signal-to-distortion ratio* (wSDR) [52] as the frequency- and time-domains, i.e.,

$$\mathcal{L}^J_{\text{MDL}} = \mathcal{L}^J_{\text{MSE}} + \alpha \left( \mathcal{L}^J_{\text{wSDR}} + 1.0 \right), \qquad (3)$$

where $\alpha$ is a scaling parameter for mixing multiple domains of loss. Specifically, $\mathcal{L}^J_{\text{MSE}}$ and $\mathcal{L}^J_{\text{wSDR}}$ are respectively calculated as follows:

$$\mathcal{L}^J_{\text{MSE}} = \sum_{j=1}^{J} \sum_{t,f} \left\{ |Y_j(t,f)| - |\hat{Y}_j(t,f)| \right\}^2, \qquad (4a)$$

$$\mathcal{L}^J_{\text{wSDR}} = \sum_{j=1}^{J} \left\{ -\rho_j \frac{y_j^{\text{T}} \hat{y}_j}{\|y_j\| \|\hat{y}_j\|} - (1-\rho_j) \frac{(x-y_j)^{\text{T}}(x-\hat{y}_j)}{\|x-y_j\| \|x-\hat{y}_j\|} \right\}, \qquad (4b)$$

where $t$ and $f$ denote the indexes of the time frame and frequency bin of the spectrogram $Y_j(t,f)$, respectively. In addition, $\rho_j$ is the energy ratio between the $j$th source $y_j$ and mixture $x$ in the time-domain, i.e., $\rho_j = \|y_j\|^2/(\|y_j\|^2 + \|x - y_j\|^2)$. Note that the output range of the wSDR in Eq. (4b) is bounded to $[-1, 1]$.

Therefore, $\left(\mathcal{L}_{\text{wSDR}}^J + 1.0\right)$ written in Eq. (3) is bounded to [0, 2.0], and it is useful to mix with another type of loss, i.e., MSE in our case. Although the SDR is traditionally calculated including the logarithm, we keep the no-logarithm style and use Eq. (4b) for MDL due to the above reason.

By using MDL, the target DNN can leverage the advantage of both domains even if the original network operates in either one of them. MDL can also be applied to many conventional DNN-based MSS methods by simply replacing the loss function; thus, no additional calculation is required during the inference.

### 3.2 Combination schemes

In this subsection, we explain bridging operation for DNN-based MSS (Section 3.2.1) and CL (Section 3.2.2) to help independent extraction networks support each other.

#### 3.2.1 Bridging operation

As shown in the blue rectangle of Fig. 2a, if DNN-based MSS is achieved using independent instrument networks, it is difficult for each network to take into account their mutual effect. Thus, we argue that it is effective to cross the network graphs to help independent sub-networks support each other[3]. This is the reason X-scheme includes bridging operation. Note that we adopt a just simple averaging layer as bridging operation. There may be some possible ways to joint sub-networks: using other techniques like cross-attention [53], squeeze-and-excitation [54], and transform-average-concatenate [55]. But we consider that they may increase the computational cost and some parameters which are supposed to be learned. One of our motivations is enhancing the existing DNN-based MSS methods keeping calculation cost and original simplicity as much as possible, and thus we focus on adding a simple averaging layer as bridging operation. Please note that the bigger size of CPU/GPU memory tends to be necessary since our X-scheme requires to put all sub-networks, each of which is used to separate an instrument, on CPU/GPU in parallel during training. But this is only a bottleneck during training and might require to adjust the batch size. When doing separation, i.e., inference, this is in general not a problem anymore due to the batch size of one.

We previously did not investigate the detailed settings of bridging operation such as its position and numbers. As shown in Fig. 2b, it is possible to place a bridge between layers #$l$ and #$(l+1)$. We can place multiple bridges depending on the number of target network layers $L$, namely, we can place up to $(L-1)$ bridges. Namely, we connect the paths to cross each source's networks by adding one or more average operators to the original network. Note that bridging operation does not have any learnable parameters; thus, the calculation cost slightly increases compared with the original network due to merely adding a few averaging operations. We can then regard the parts before and after the last added bridge as the interaction and each source extraction part; thus, their capacity depending on the position of bridging affects the final MSS performance. Motivated by the above discussion, we will conduct experiments on X-UMX (Section 4.3) to confirm the effect of the number and position of bridging operation on MSS performance.

#### 3.2.2 Combination loss

As mentioned above, the purpose of applying bridging operation is to enable each source-extraction network to handle the relationship among output sources via built bridges. In other words, it is necessary for each source-extraction network to learn its mutual relationships during training. However, using only bridging operation is insufficient for the networks to work together if the loss function is computed independently for each instrument. Thus, it is effective to cross the loss function as well as network paths via CL to boost the benefit of the built bridges. For CL, we consider the combinations of output spectrograms to enable each DNN-based source extractor to interact with each other. Specifically, we combine two or more estimated spectrograms into new ones, where each one can extract two or more sources from the mixture. Using the newly obtained combination spectrograms enables us to compute more loss functions than when we use only the individual instrument spectrograms independently, i.e.,

$$\mathcal{L} = \frac{1}{N}\sum_{n=1}^{N} \mathcal{L}_{\text{MDL}}^n, \tag{5}$$

where $N > J$ is the total number of possible combinations except for mixing all $J$ sources, i.e., $N = \sum_{i=1}^{J-1} \binom{J}{i}$, and $n$ denotes the index of the $n$th combination[4]. For instance, when separating $J = 4$ sources, as is the case with MUSDB18, we can consider 14 ($= \sum_{i=1}^{4-1} \binom{4}{i}$)

---

[3] Note that bridging operation may be only needed for methods such as UMX, since it consists of individual extraction networks. In other words, this bridging is not necessary for methods that learn one network for all sources such as Demucs [39].

[4] Initial experiments showed that the combination $_JC_J$, i.e., adding and mixing all sources, does not further improve MSS performance, thus it is not used in Eq. (5).

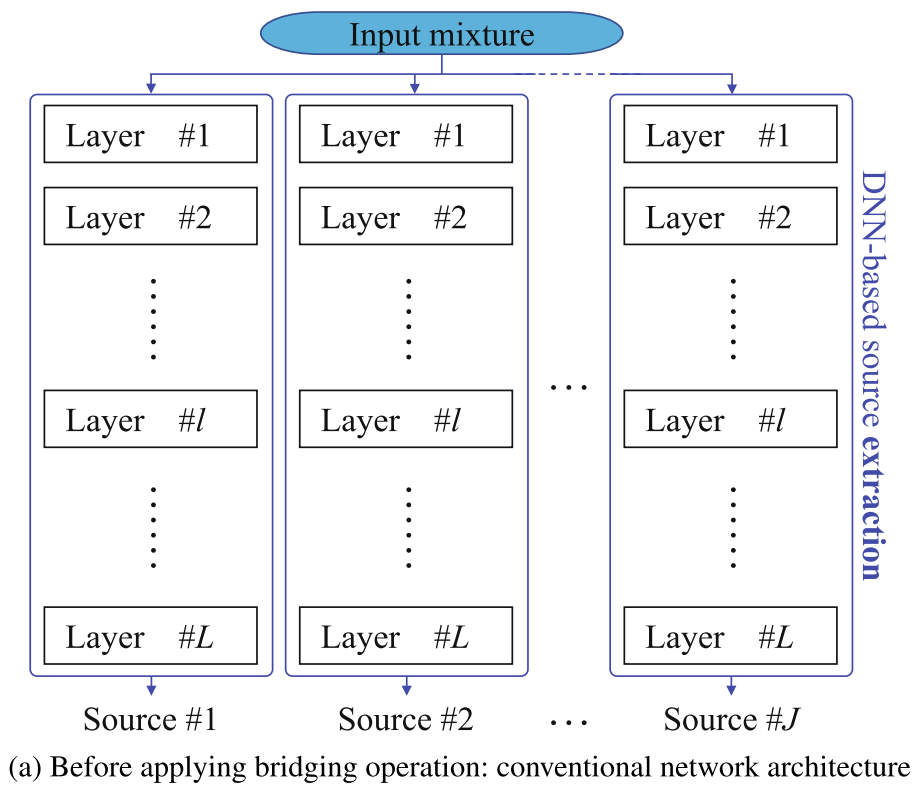

(a) Before applying bridging operation: conventional network architecture

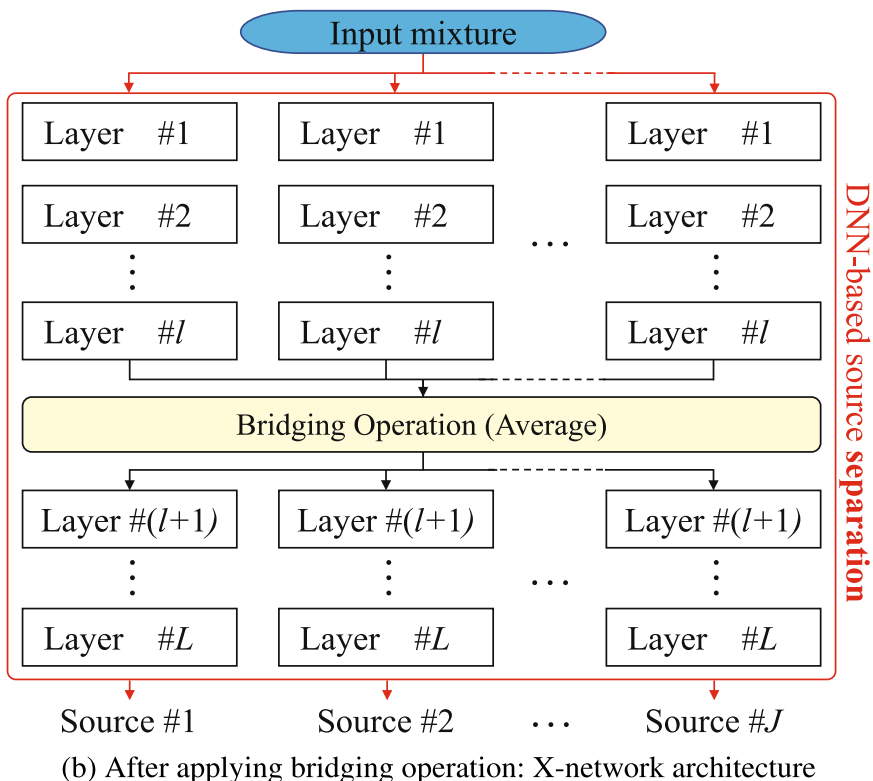

(b) After applying bridging operation: X-network architecture

**Fig. 2** Comparison of applying and not applying bridging operation

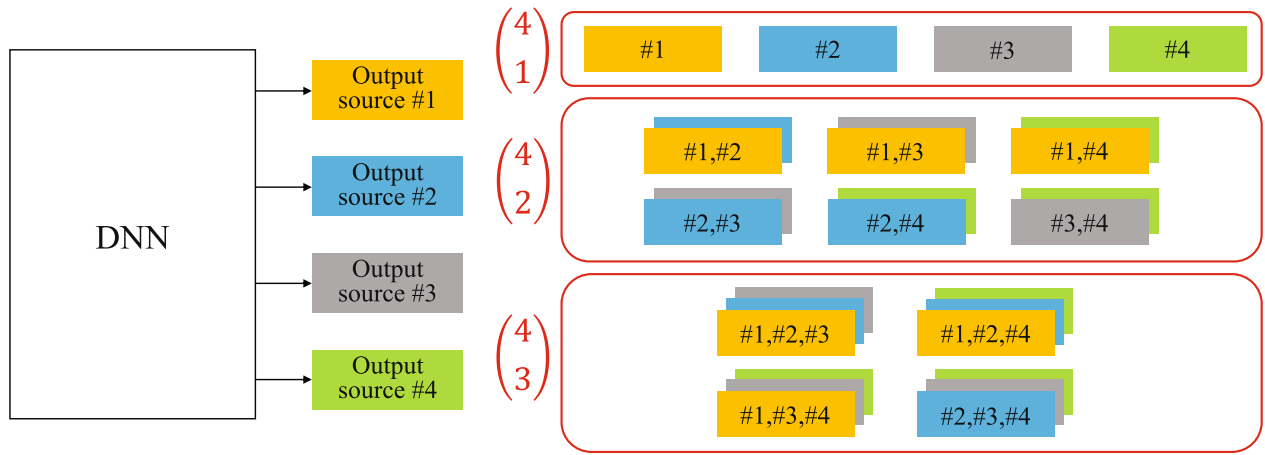

**Fig. 3** CL when mixture consists of four sources

combinations in total, as shown in Fig. 3, whereas conventional methods handle only each source independently, i.e., 4 source spectrograms.

To explain the advantage of CL, let us consider the following example. Assume that we have a system with leakage of *vocals* into *drums* and *bass* resulting in similar errors that both instruments exhibit. By considering the combination *drums + bass*, we notice that the two errors are correlated, resulting in an even larger leakage of vocals, which we try to mitigate using CL. More formally, let $\epsilon_j$ denote the prediction error of the *j*th source; $\hat{y}_j = y_j + \epsilon_j$. We can then consider the MSE of the combination $u = y_1 + y_2$:

$$\begin{aligned}\text{MSE}(u, \hat{u}) &= \mathbb{E}\left[(u - \hat{u})^2\right] \\ &= \mathbb{E}\left[\{(y_1 + y_2) - (y_1 + \epsilon_1 + y_2 + \epsilon_2)\}^2\right] \\ &= \mathbb{E}\left[\epsilon_1^2 + 2\epsilon_1\epsilon_2 + \epsilon_2^2\right].\end{aligned}$$

When we consider $y_1$ and $y_2$ separately without the combination, the term "$2\epsilon_1\epsilon_2$" does not appear in the MSE; $\text{MSE}(y_1, \hat{y}_1) + \text{MSE}(y_2, \hat{y}_2) = \mathbb{E}\left[\epsilon_1^2 + \epsilon_2^2\right]$. Therefore, by using CL, we can monitor the error correlation term "$\mathbb{E}[2\epsilon_1\epsilon_2]$," which helps the source-extraction networks train when they are correlated. Specifically, we expect the term "$\mathbb{E}[2\epsilon_1\epsilon_2]$" to be able to detect errors leaking into the wrong track. In order to efficiently reduce this term, we use the bridging operations which allows each sub-network to be aware of the others and, hence, to reduce potential leakage to a wrong source. Specifically, tying networks together helps the training as now also gradient information is exchanged which can help to learn to have a small "$2\epsilon_1\epsilon_2$" term. Furthermore, they also benefit from a joint feature extraction. Therefore, we bridged the network by just adding simple average operators as shown in Fig. 2b, which turned out to be beneficial since their results were actually improved in spite of using the same configurations except applying our X-scheme.

We can also analyze CL in terms of a geometrical viewpoint. Focusing on Eq. (4b), since the wSDR consists of two cosine similarity functions, it monitors the angle consisting of the ground truth $y_j$ and corresponding predicted $\hat{y}_j$. However, there is a critical case in which the prediction error cannot be detected in terms of the cosine similarity. As shown in Fig. 4, when the predicted $\hat{y}_1$ and $\hat{y}_2$ are respectively orthogonal to the corresponding ground truth $y_1$ and $y_2$, it is difficult to detect the prediction error since $\cos(y_1, \hat{y}_1)$ and $\cos(y_2, \hat{y}_2)$ are both zeros. However, CL can detect its prediction error via the combined signals $u$ and $\hat{u}$ since the score of $\cos(u, \hat{u}) = -1$ penalizes the target network by substituting it for the wSDR-based loss function. There is possibly a case that all of $\cos(u, \hat{u})$, $\cos(y_1, \hat{y}_1)$, and $\cos(y_2, \hat{y}_2)$ simultaneously become zero. However, in such case, all vectors (including their sums) are orthogonal, CL just does not bring a benefit but also does not cause any degradation. Namely, there is no harm and it is just not effective. Furthermore, if we would include the multichannel Wiener filter (MWF) like UMX and X-UMX, then we can expect that this case can not appear as MWF redistribute the residual to all sources and by this always have a non-orthogonal sum which results in an error. Note that we need to apply our X-scheme after using MWF in that case.

Independent sub-networks can detect each other via the added bridges and CL. The DNN-based MSS network extended with X-scheme can handle multiple sources together, i.e., separate two or more sources, rather than each source independently. From a different viewpoint, CL can be considered to provide a similar benefit to multi-task learning [56] since it handles multiple objectives jointly by computing combinational loss functions.

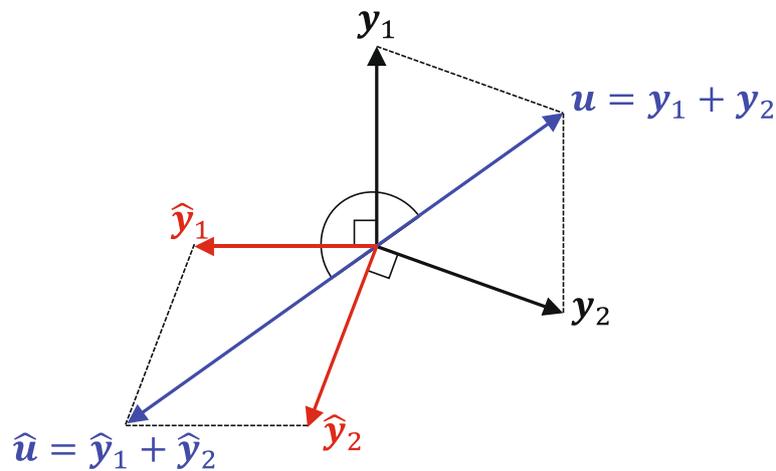

**Fig. 4** Example of angles consisting of ground truth signals ($y_1, y_2, u$) and their corresponding predicted ($\hat{y}_1, \hat{y}_2, \hat{u}$). Black and red arrows denote ground truth and corresponding predicted signals, respectively. Blue arrows denote combined signals

We can apply X-scheme to many DNN-based MSS methods to improve their performances while maintaining almost the same computational cost as the original method since MDL and CL are merely loss functions and bridging operation is achieved with simple average operations without increasing learnable parameters. As discussed in Section 4, X-scheme improves DNN-based MSS performance.

## 4 Experiments

In this section, we present our experiments on X-scheme for MSS. We first explore the effect of the bridging position using X-UMX [30] to provide insights on the optimal position and its sensitivity. Next, we confirm the scalability of X-scheme in a large-scale data regime. Finally, we demonstrate the generality of X-scheme by applying it to another type of network architectures, D3Net and Conv-TasNet.

We used the following datasets and STFT/ISTFT settings for the experiments.

### 4.1 MUSDB18 [57]

The MUSDB18 dataset is comprised of 150 songs, each of which was recorded at a 44.1-kHz sampling rate. It consists of two subsets ("train" and "test"), where we further split the train set into "train" and "valid" as defined in the official "musdb" package[5]. For each song, the mixture and its four sources, i.e., *bass*, *drums*, *other*, and *vocals*, are available.

### 4.2 STFT/ISTFT

We used a Hann window with a length of 4096 samples and 75% overlap. We used STFT magnitudes obtained from the mixture signal as input and trained networks to estimate target mask $M_j(t,f)$ or spectrograms $Y_j(t,f)$, where $f$ is the frequency bin and $t$ the frame index. To use STFT and ISTFT as differentiable layers for MDL, we used "torch.stft" and "torch.istft" from PyTorch which are readily available and provide a differentiable implementation of the STFT/ISTFT[6]. Please see also https://github.com/asteroid-team/asteroid for our actual implementation.

### 4.3 X-UMX

The network architecture of UMX is illustrated in Fig. 5a. The network was trained to estimate all the sources' masks with the Adam [58] optimizer for 1000 epochs. The learning rate was set to 0.001 with a weight decay of 0.00001. The batch size was set to 14 and each input was a random crop of 6.0 sec from the dataset. The scaling parameter $\alpha$, introduced in Eq. (3) for MDL was set to 10.0 to approximately equalize the ranges of $\mathcal{L}^J_{\mathrm{MSE}}$ and $\mathcal{L}^J_{\mathrm{wSDR}}$ by looking at each loss function's learning curves, respectively. Note that the details of other settings are shown in our code[7] and previous paper [30].

#### 4.3.1 Bridging positions

As shown in Fig. 2, bridging operation can be applied to arbitrary positions between the layers. The number of bridges can also be increased depending on the number of gaps between adjacent layers. Therefore, in this section, we present the results regarding the position of bridging operation on X-UMX trained under the same configurations, e.g., the number of epochs, regularization parameters, and type of optimizer, as mentioned in the previous subsection. We show the simplified network architecture and possible bridging candidates of UMX in Fig. 5b. UMX roughly consists of three affine blocks and a BLSTM block. Each affine block has a fully connected layer, batch normalization layer, and activation function. The BLSTM block has three consecutive BLSTM layers with dropout. In this experiment, we considered three positions as candidates for inserting the bridging network and examined the performance for all combinations of bridging position.

---

[5] https://github.com/sigsep/sigsep-mus-db/blob/master/musdb/configs/mus.yaml

[6] Some famous DNN libraries such as PyTorch and TensorFlow already provided official implemented STFT and ISTFT layers.

[7] https://github.com/asteroid-team/asteroid/tree/master/egs/musdb18/X-UMX

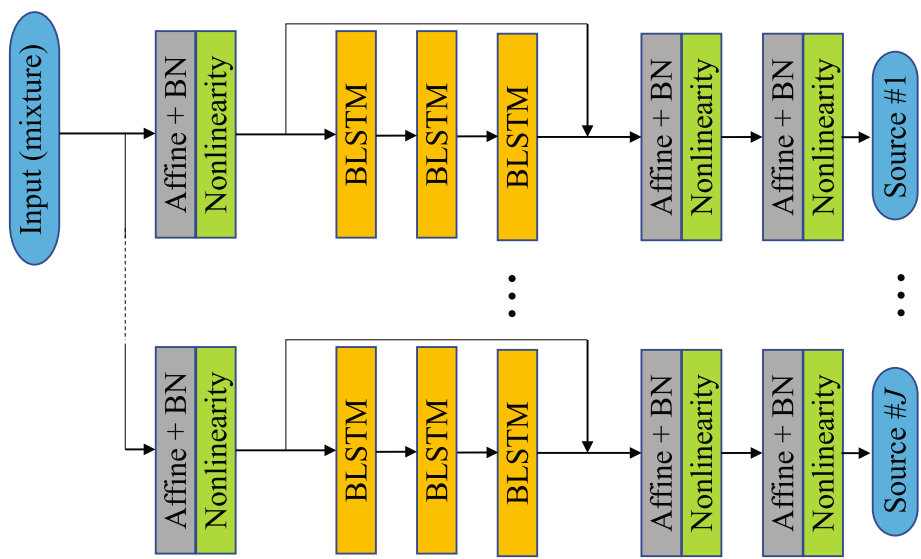

(a) UMX

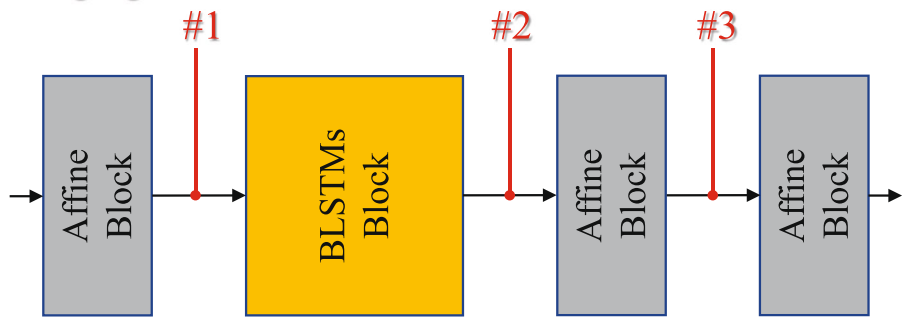

(b) Candidates of bridging operation on UMX

**Fig. 5** Original network architecture of UMX and its bridging candidates

The results are shown in Fig. 6. The performances of almost all bridged versions of UMX, i.e., bridging position (BP) from 1 to 7 (BP1-BP7), were superior to the baseline from the perspective of source-to-interference ratio (SIR) and source image-to-spatial distortion ratio

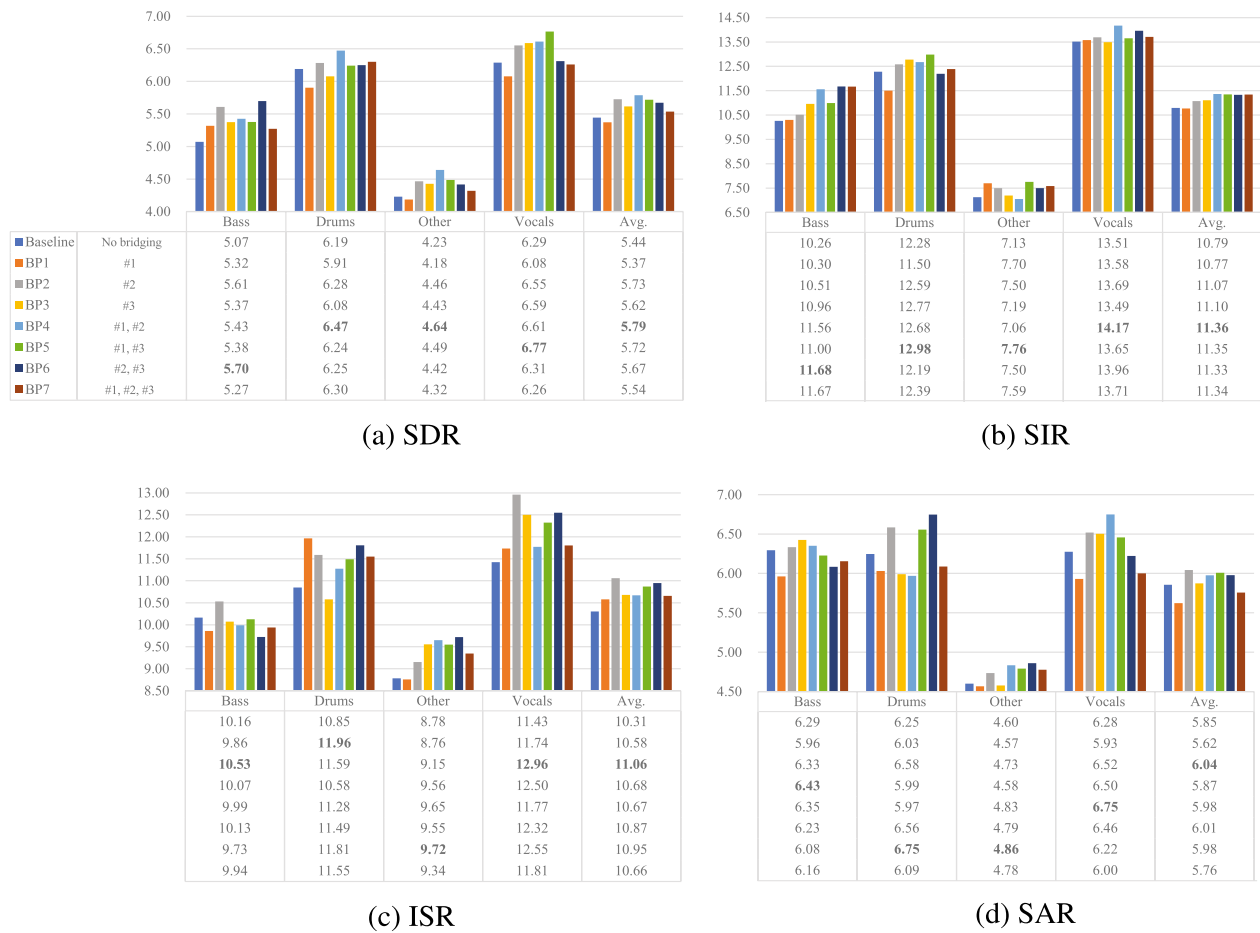

**Fig. 6** Results regarding bridging position(s) for X-UMX. Note that correspondence between bridging indexes written in **a** and positions are shown in Fig. 5b

(ISR) (see "Avg." of Fig. 6b and c). Only the SIR result of BP1 did not outperform that of the baseline but was comparable. Hence, we argue that bridging operation can improve the suppression of the other interference instruments without increasing linear distortions since ISR becomes low when the output signal increases linear distortions. Focusing on the SDR results, which were computed by summing up the weighted SIR, ISR, and SAR, we argue that X-UMX outperformed UMX because the SDR results of BP1-BP7 improved in most cases compared with that of the baseline (see Fig. 6a). In particular, BP4, which bridged the paths between "Affine Block" and "BLSTMs Block," performed the best in terms of the SDR. Hence, bridging paths between the gaps of different type of blocks or layers is probably effective in terms of sharing each sub-extraction network's information.

### 4.3.2 Effectiveness of CL

First of all, we confirmed the validity of the term "$2\epsilon_1\epsilon_2$" mentioned in Section 3.2.2, which is ignored in the case of training each of instruments' sub-networks separately. Not only this term but also bridging networks bring benefit. Tying networks together helps the training as now also gradient information is exchanged which can help to learn to have a small "$2\epsilon_1\epsilon_2$" term. Furthermore, they also benefit from a joint feature extraction. In this way, by computing this term through our X-scheme, we take this mutual effect among sources into account when training the DNN. Specifically, by considering this term in the loss function, it is expected to penalize an errors having correlation between instruments when either of an instrument is wrongly separated to the wrong track. To confirm this, we compared the actual separated results of UMX and X-UMX. As shown in Fig. 7, X-UMX succeeded to suppress errors in the vocals track leaked from drums as expected. In particular, the regions highlighted by colored rectangles were obvious, and this improvement was also audible. It is considered that this power, i.e., energy from drums, which leaked in the wrong track was penalized through the loss function by our X-scheme as we discussed in Section 3.2.2 resulting in the performance improvement shown in Fig. 7a.

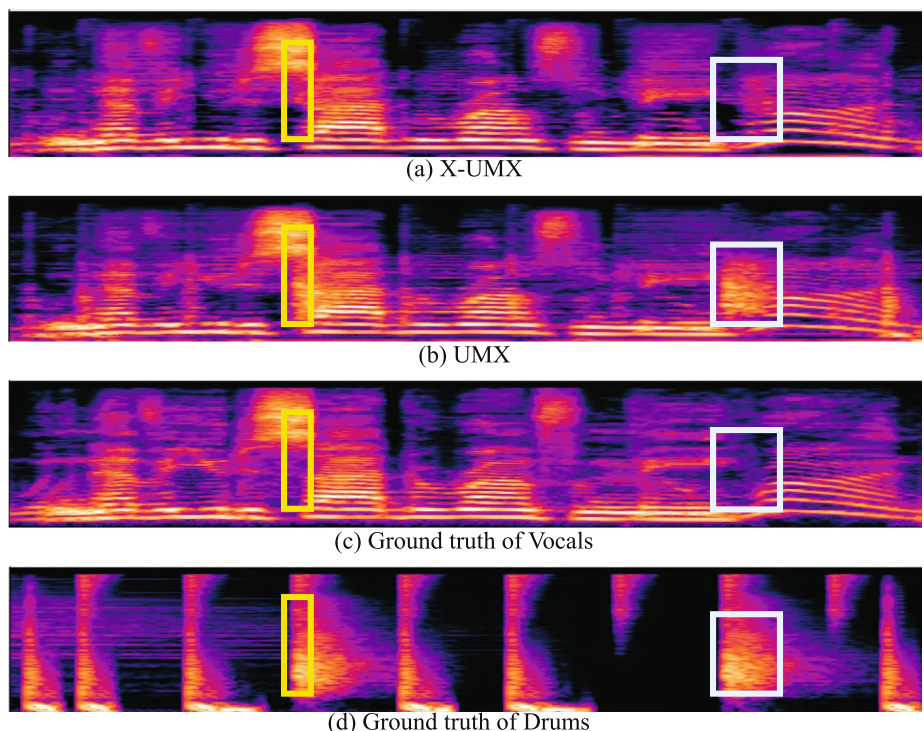

**Fig. 7** An example of comparing the synchronized spectrograms of vocals obtained by applying X-UMX and UMX to a musical piece of MUSDB18 test set. The rectangles depicted in the same color denote the corresponding time-frequency regions with each other. For sake of simplicity, only single channels were shown here although the actual results were stereo

To confirm the validity of CL in more detail, we monitored the performance change according to the number of combinations. As we discussed in Section 3.2.2, the combined vector may not potentially bring a benefit especially when the number of the target sources is few. Therefore, we fixed the target source as "vocals" and trained 2 X-UMXs each of which was respectively trained by using 3 and 4 instruments for CL and bridging operation. Note that both of them always received the mixture signal consisting of 4 instruments as input, but the number of output instruments, i.e., sub-networks, was different. Namely, "X-UMX on 3 sources" separated 3 instruments from input that was 4 sources mixture whereas "X-UMX on 4 sources" separated 4, i.e., full, instruments from input. Then, in terms of CL, the number of combinations used for CL was different. The results are summarized in Table 1.

As shown in the table, all results of "X-UMXs on 3 sources" were inferior to the model with all four sources. Intuitively, the more related to the vocals the excluded source was, the worse performance the results tended to be. The power of "Bass" is concentrated on lower frequency bands, and thus "Bass" has lower correlation with "Vocals" than "Drums."

**Table 1** Performance comparison of separating "Vocals" track of MUSDB18 test set on X-UMXs having the different number of output instruments in SDR [dB]. Since our X-scheme uses the combination of separated output instruments, the number of combinations changes and affects performance when the number of output instruments is different. Note that we removed the post-processing, multichannel Wiener filter (MWF) [19] which was originally used in X-UMX [30], in this experiments since it requires all separated instruments

|  | X-UMX on 3 sources | | X-UMX on 4 sources |
| --- | --- | --- | --- |
|  | w/o bass | w/o drums |  |
| Vocals | 6.01 | 5.97 | **6.21** |

Table 2 Comparison of X-UMX with UMX in terms of SDR("median of frames, median of tracks"). Note that all results were evaluated on MUSDB18 test set

| Method | Training dataset size | Bass | Drums | Other | Vocals | Avg. |
| --- | --- | --- | --- | --- | --- | --- |
| UMX [24] | 10h (MUSDB18) | 5.23 | 5.73 | 4.02 | 6.32 | 5.33 |
| X-UMX |  | **5.43** | **6.47** | **4.64** | **6.61** | **5.79** |
| UMXL | 100h (INTERNAL) | 5.79 | 6.93 | 4.50 | 6.71 | 5.98 |
| X-UMXL |  | **6.28** | **7.39** | **4.83** | **7.57** | **6.52** |
| Public UMXL[a] | 500h | 6.02 | 7.15 | 4.89 | 7.21 | 6.32 |

[a] https://zenodo.org/record/5069601

### 4.3.3 Scalability with large training datasets

In this section, we discuss the potential of X-UMX for a large-scale training dataset, i.e., X-UMXL, which was not assessed in our previous study [30]. DNNs can generalize well if enough data is available for training, and some regularization methods might become ineffective in such a case. Thus, it is important to investigate the scalability of X-scheme. Specifically, we trained UMX and X-UMX on an internal dataset consisting of 1505 songs with a total duration of approximately 100 h, which is 10 times larger than MUSDB18. The dataset exhibits a diverse linguistic composition, with 63% of the songs being in English, 20% in French, 6% in German, and the remaining 11% comprising various other languages such as Italian, Spanish, and Dutch. Regarding musical genres, the collection predominantly features pop and rock music. It also includes a selection of country songs and movie soundtracks, though these are less prevalent. We denote this dataset as "INTERNAL" and note that it has no overlapped songs with MUSDB18. Each song of INTERNAL consists of four instruments, as in MUSDB18.

The results are summarized in Table 2. X-UMX and X-UMXL outperformed the corresponding UMX and UMXL if they were trained on the same dataset, i.e., using MUSDB18 or INTERNAL. X-UMX and X-UMXL outperformed the original UMX and UMXL for all instruments (see the boldface in Table 2). This shows that X-scheme is effective even when we have more training data available.

It is worth noting that X-UMXL greatly outperformed not only our self-implemented UMXL trained on INTERNAL but also "public UMXL," which was provided by the authors of UMX, although the size of our dataset is one fifth of theirs[8] (see the yellow highlighted cells in Table 2). From these results, we argue that X-scheme can use a given dataset for training more successfully, and even outperform a traditional setup with more training data.

### 4.3.4 X-D3Net and X-Conv-TasNet

Next, we firstly integrated X-scheme into D3Net resulting in X-D3Net. The network architecture is shown in Fig. 8. The original D3Net, C1, uses band-wise MDenseNets [21] and integrated their outputs by applying a dense block, but they are independent of each other, i.e., there is no path to share their relationship among them. Hence, the bridging path is added at the end of band-wise D3 blocks, resulting in X-D3Net, as in the experiment in Section 4.3.1. This suggests that the semantic boundary can be a good position for inserting bridging operation. The differences in network the architecture between D3Net and X-D3Net are shown in Fig. 8. Each network of X-D3Net was trained to estimate all the sources' spectrograms with the Adam [58] optimizer for 70 epochs. The initial learning rate was set to 0.001 with a weight decay of 0.00001, and its learning rate was dropped to 0.0003 and 0.0001 after 40 and 60 epochs, respectively. The batch size was set to 6 and each input was a randomly cropped music spectrogram with 352 frames. The scaling parameter $\alpha$ was set again to 10, as we did for X-UMX.

The results are shown in Fig. 9. Note that "P" denotes the proposed method that includes all components of X-scheme, i.e., MDL, bridging operation, and CL, while "C1-C7" denote the comparative methods lacking some of these components in order to confirm their effectiveness one-by-one. In terms of the SDR, the methods using at least one component of X-scheme, i.e., C2-C7 and P, were superior to D3Net, i.e., C1 (see the average performances denoted as "Avg." in Fig. 9a). Therefore, the validity of each component of X-scheme was confirmed on a CNN-based MSS method (D3Net) as well as an RNN-based MSS method (UMX). Overall, we could improve MSS performance by 0.3 dB.

In particular, the positive effect of MDL was notable compared with our previous corresponding results on X-UMX [30] (see the results of methods including MDL, i.e., C2, C5, C6, and P). Therefore, regardless of whether the target network is an originally integrated one or tied against independent source sub-networks, the loss-function-related core components of X-scheme, i.e., MDL and CL, can improve MSS performance.

---
[8] The size of dataset used for training UMXL, i.e., 500 h, was confirmed with the developers.

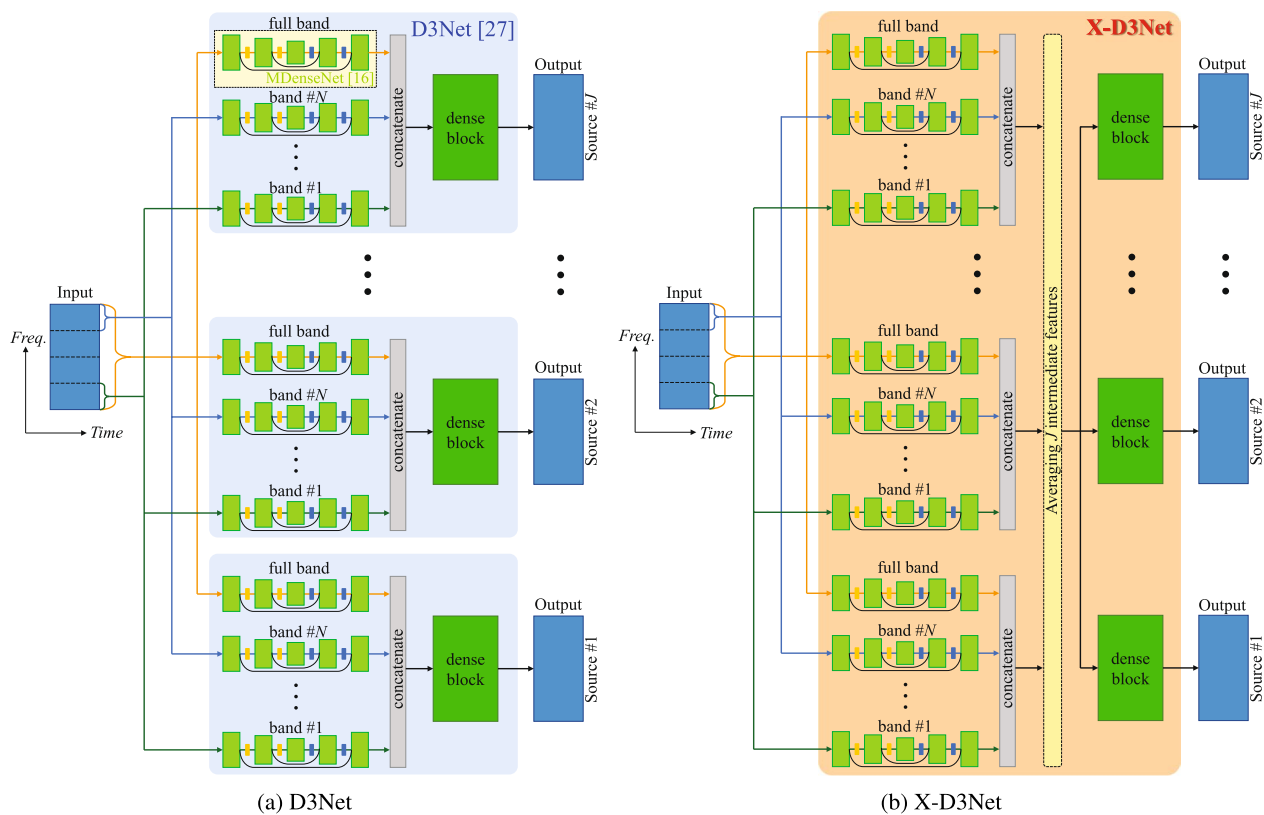

**Fig. 8** Comparison of D3Net and X-D3Net

Finally, to see the effectiveness of applying our X-scheme to DNN-based MSS methods, we summarize the performance comparison before and after applying X-scheme in Table 3. To confirm the effectiveness for not only frequency-domain network, i.e., UMX and D3Net used in the above, but also time-domain network, we also applied X-scheme to Conv-TasNet [1] resulting in X-Conv-TasNet. Note that time-domain networks tend to require much larger size of memory and the corresponding training time than frequency-domain ones, and thus we did not run an ablation study but instead applied our X-scheme to Conv-TasNet using the learnings from X-UMX and X-D3Net. As shown in Table 3, all instruments were improved when comparing to the vanilla networks. In addition to the quantitative results, we also studied their spectrograms shown in Fig. 10. As shown in the figure, all vanilla methods tend to miss the power of "Other" track, but all of them became to be able to detect and extract it by applying our X-scheme. This is due to the missed power that leaked in wrong tracks which was penalized through the X-scheme loss function as we discussed in Section 3.2.2. Thus, we can argue that our X-scheme works well not only for frequency-domain networks but also for time-domain ones, e.g., Conv-TasNet.

From the aforementioned results, we can conclude that our X-scheme can be applied to diverse types of networks such as CNN-based time- and frequency-domain models as well as RNN-based time- and frequency-domain ones. However, please note that the detailed effect of our method, e.g., where the most effective bridging position is, the number of combinations and bridges that should be used, and what types of time-domain and frequency-domain loss functions are effective, may be different depending on the detailed characteristics of the target network. Therefore, it is important to insert each core part of X-scheme, (i) MDL, (ii) bridging operation, and (iii) CL, one-by-one and adapting them such that the optimal configuration for the target network is found.

## 5 Conclusion

We revisited our previous proposal and summarized its core component, a versatile scheme called X-scheme. X-scheme consists of three parts: (i) MDL, (ii) bridging operation, and (iii) CL, which improve the performance of DNN-based MSS with almost no increase in calculation cost. Specifically, as MDL and CL are merely loss functions used during training, they do not affect the computational cost at inference. As shown in Fig. 2, bridging operation does not increase calculation cost due to adding only a few average computations. To verify X-scheme for another type of network that differs from the recurrent type, i.e., UMX, we derived an X-scheme-based convolutional networks in this paper. The frequency-domain and time-domain convolutional networks extended by X-scheme are respectively X-D3Net and X-Conv-TasNet. We confirmed their validity compared to the original ones through experiments. We also examined the detailed effectiveness of X-UMX through the following experiments: (a) searching for the effective bridging position(s) and b) using a larger dataset than MUSDB18. X-UMXL, obtained by training X-UMX on our large-scale data regime (INTERNAL), greatly outperformed the public UMXL, although the size of our dataset is about 20% that of the private training dataset for public UMXL. Therefore, by sharing the information regarding the progress of MSS among all sub-networks, our X-scheme-based MSS method can versatilely improve MSS performance more effectively than the original method. It is worth noting that X-scheme has high practicability since its components, i.e., MDL, CL, and bridging operation, have almost no effect on the inference of the original target method and are easy to implement.

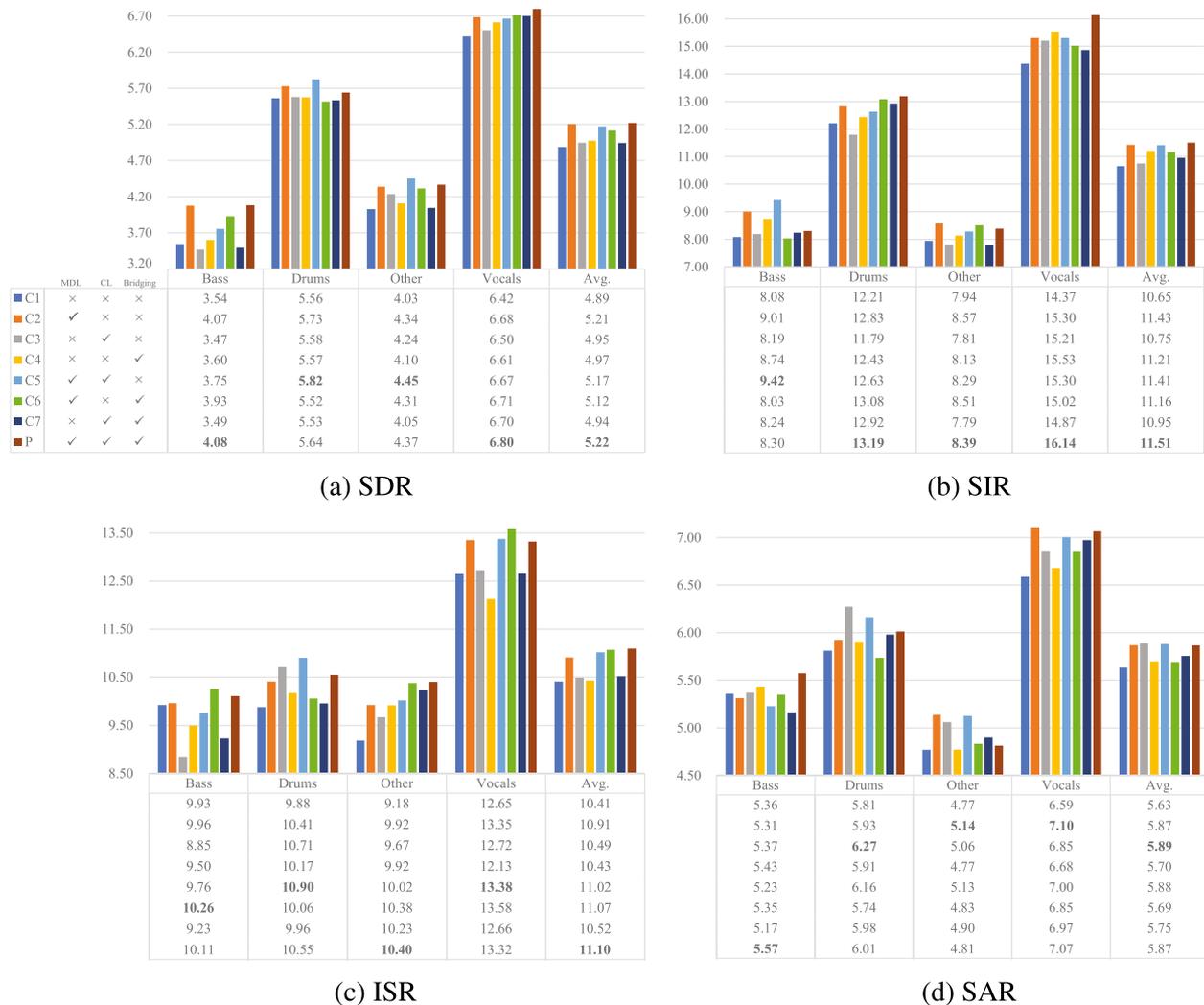

**Fig. 9** Experimental results of X-D3Net. Note that conventional D3Net, which is denoted as "C1," was re-trained for this experiment to equalize number of optimizers with X-D3Net's. Namely, each version of X-D3Net was trained on single optimizer while original paper trained four D3Nets, each of which corresponds to each instrument, using four optimizers separately. Thus, results of C1 are different from those of original paper

**Table 3** Comparison of our X-scheme by applying to 3 different MSS methods in terms of SDR ("median of frames, median of tracks"). Note that all results were trained on our local environment by using each author's official codes

|  | Bass | Drums | Other | Vocals | Avg. |
|---|---|---|---|---|---|
| UMX [24] | 4.93 | 5.72 | 4.00 | 6.09 | 5.18 |
| X-UMX | **5.43** | **6.47** | **4.64** | **6.61** | **5.79** |
| D3Net [34, 35] | 3.54 | 5.56 | 4.03 | 6.42 | 4.89 |
| X-D3Net | **4.08** | **5.64** | **4.37** | **6.80** | **5.22** |
| Conv-TasNet [1] | 6.23 | 5.60 | 3.97 | 6.90 | 5.66 |
| X-Conv-TasNet | **6.29** | **6.57** | **4.69** | **7.19** | **6.18** |

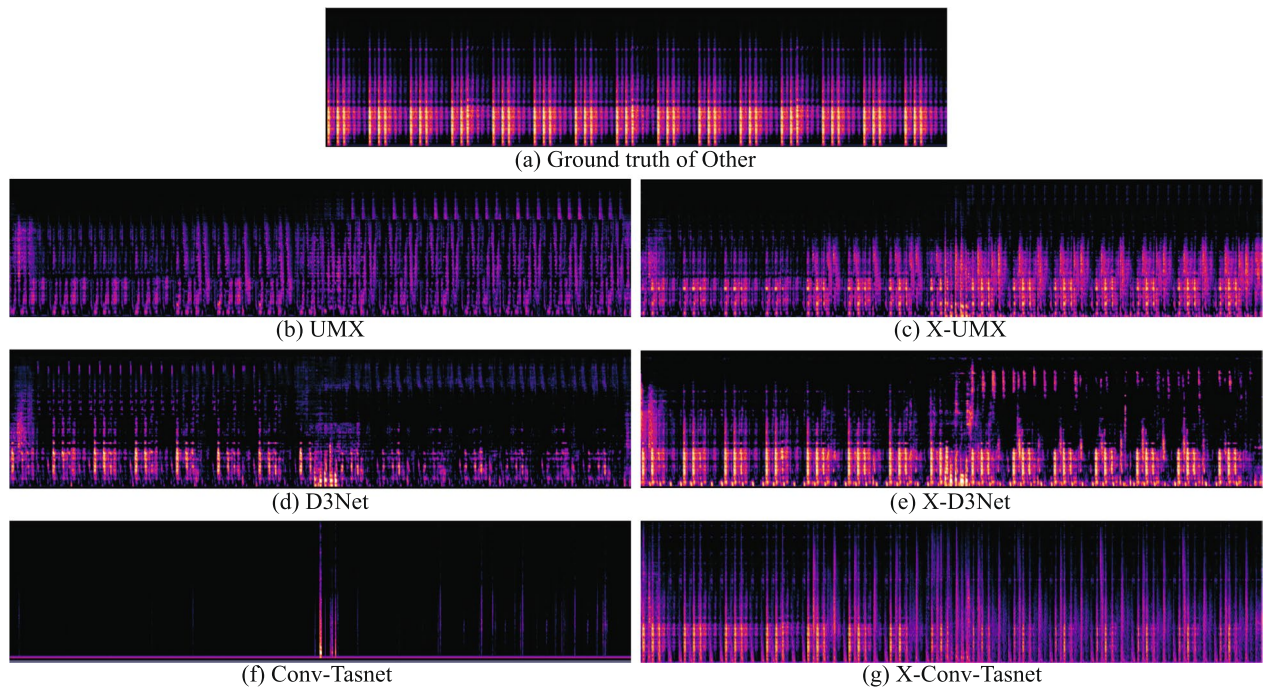

**Fig. 10** Comparison between before and after applying X-scheme to UMX, D3Net, and Conv-TasNet on an example of MUSDB18 test set. Note that all spectrograms are synchronized. For sake of simplicity, only single channels were shown here although the actual results were stereo

## Appendix 1: Robustness against initialization

To confirm the robustness against the initial random seed, we conducted experiments that ran X-UMX on the same experimental settings except the initialization. Specifically, we ran 10 X-UMXs where each was initialized with different random seeds and compared the final performances among them. As shown in Table 4, all instruments' averaged results among all different seeds outperformed the vanilla UMX's ones. Furthermore, even by focusing each random seed's results one-by-one, there are almost no scores that were inferior to the UMX's ones. Therefore, we argue that our X-scheme enables the target network to enhance its performance independent of initialization effects.

**Table 4** Experimental results of X-UMXs each of which was initialized from 10 different random seeds in SDR [dB]

| | Random seed | | | | | | | | | | Mean± Std. | UMX [24] |
|---|---|---|---|---|---|---|---|---|---|---|---|---|
| | #1 | #2 | #3 | #4 | #5 | #6 | #7 | #8 | #9 | #10 | | |
| Vocals | 6.53 | 6.08 | 6.67 | 6.33 | 6.17 | 6.54 | 6.64 | 6.27 | 6.41 | 6.60 | **6.42** ± 0.203 | 6.32 |
| Drums | 6.18 | 6.38 | 6.33 | 6.14 | 6.46 | 6.40 | 6.11 | 6.00 | 6.32 | 6.28 | **6.26** ± 0.147 | 5.73 |
| Bass | 5.50 | 5.56 | 5.39 | 5.45 | 5.32 | 5.52 | 5.39 | 5.28 | 5.61 | 5.52 | **5.45** ± 0.108 | 5.23 |
| Other | 4.48 | 4.23 | 4.53 | 4.39 | 4.29 | 4.41 | 4.43 | 4.41 | 4.49 | 4.45 | **4.41** ± 0.093 | 4.02 |

**Acknowledgements**
Not applicable.

**Authors' contributions**
RS conducted all experiments and analyzed the results. And RS was also a major contributor in writing the manuscript. NT and SU technically supervised RS and polished the initial written manuscript. All authors read and approved the final manuscript.

**Authors' information**
RS received his B.S. and M.S. degrees in Electronics and Information Engineering from Hokkaido University, Japan in 2014 and 2016, respectively. From 2022 to 2023, he had worked at the Stanford Vision and Learning Lab (SVL) at Stanford University, USA. He is currently a researcher at Sony Research and Ph.D. candidate at Graduate School of Information Science and Technology, Hokkaido University. His research interests include biosignal processing, music information retrieval, acoustic signal processing, and 3D computer vision. He is a member of IEEE.

NT received Ph.D. from University of Tsukuba, Japan, in 2020. Formerly, he had worked at the Computer Vision Lab at ETH Zurich, Switzerland. Since he joined Sony in 2008, he has performed research in the field of audio, computer vision, and machine learning. In 2018, he won the Sony Outstanding Engineer Award, which is the highest form of individual recognition for Sony Group engineers. He achieved the best scores in several challenges including Signal Separation Evaluation Campaign (SiSEC) 2018 and Detection and Classification of Acoustic Scenes and Events (DCASE) 2021. He has authored several papers and served as a reviewer at CVPR, ICASSP, Interspeech, ICCV, Trans. ASLP, Trans. MM, and more. He co-organized the DCASE 2022 task3.

SU received the Dipl.-Ing. and Ph.D. degree in electrical engineering from the University of Stuttgart, Germany, in 2006 and 2012, respectively. From 2007 to 2011, he was a research assistant at the Chair of System Theory and Signal Processing, University of Stuttgart. In this time, he worked in the area of statistical signal processing, focusing especially on parameter estimation theory and methods. Since 2011, he is with the Sony Stuttgart Technology Center where he works as a Senior Principal Engineer on problems in music source separation and deep neural network compactization.

ST received his B.S. degree in communications engineering and M.S. degree in information science from Tohoku University, Japan, in 2000 and 2002, respectively. He is currently a researcher and a senior manager at Sony Group Corporation, Japan. His research interests include speech and audio signal processing and machine learning. His team achieved the first place in DCASE 2021 Challenge in Task3 and co-organized DCASE Challenge in 2022 and 2023. He also co-organized Sound Demixing Challenge 2023, where real audio from movies by Sony Pictures Entertainment were used for the evaluation.

YM received the B.S. and M.S. degrees in information science from Keio University in 2002 and 2004, respectively. He obtained the Ph.D. degree in information science and technology from the University of Tokyo in 2020. Currently, he is leading Creative AI Lab at Sony Group Corporation while serving as Specially Appointed Associate Professor at Tokyo Institute of Technology. He joined Sony Corporation in 2004 and has been leading teams that developed the sound design for the PlayStation game title called "Gran Turismo Sport" and spatial audio solution called "Sonic Surf VR." He also won several awards such as TIGA award for best audio design for Gran Turismo Sports and a jury selection at Japan Media Arts Festival for their 576-channel sound field synthesis called "Acoustic Vessel Odyssey." From 2011 to 2012, he was a visiting researcher at Analysis/Synthesis Team, Institut de Rechereche et Coordination Acoustique/Musique (IRCAM), Paris, France. He was involved in the 3DTV content search project sponsored by European Project FP7, in research collaboration with IRCAM. In 2021, his team organized Music Demixing (MDX) Challenge where Sony Music provided a professionally-produced music dataset for the evaluation of submitted systems to an online platform on Alcrowd. His team also participated in DCASE2021 Challenge and achieved the first place in Task3.

**Funding**
Not applicable.

**Availability of data and materials**
The dataset generated and/or analyzed during the current study are available in the "MUSDB18" repository, https://sigsep.github.io/datasets/musdb.html [57]. The dataset, "INTERNAL" (see Table 2), generated and/or analyzed during the current study are not publicly available since it is compensation and thus not public dataset.

## Declarations

**Competing interests**
The authors declare that they have no competing interests.

## References

1. Y. Luo, N. Mesgarani, Conv-TasNet: Surpassing ideal time-frequency magnitude masking for speech separation. IEEE/ACM Trans. Audio Speech Lang. Process. **27**(8), 1256–1266 (2019)
2. A.V. Oppenheim, R.W. Schafer, J.R. Buck, *Discrete-Time Signal Processing*, 2nd edn. (Prentice-hall Englewood Cliffs, USA, 1999)
3. G. Meseguer-Brocal, G. Peeters, in *Proc. of the 20th International Society for Music Information Retrieval Conference (ISMIR)*, ed. by A. Flexer, G. Peeters, J. Urbano, A. Volk. Conditioned-U-Net: introducing a control mechanism in the U-Net for multiple source separations (2019). pp. 159–165. http://archives.ismir.net/ismir2019/paper/000017.pdf. Accessed 29 Apr 2024
4. O. Slizovskaia, G. Haro, E. Gómez, Conditioned source separation for musical instrument performances. IEEE/ACM Trans. Audio Speech Lang. Process. **29**, 2083–2095 (2021). https://doi.org/10.1109/TASLP.2021.3082331
5. V.S. Kadandale, J.F. Montesinos, G. Haro, E. Gómez, in *Proc. of IEEE 22nd International Workshop on Multimedia Signal Processing (MMSP)*. Multichannel U-Net for music source separation (2020), pp. 1–6. https://doi.org/10.1109/MMSP48831.2020.9287108
6. E. Perez, F. Strub, H. de Vries, V. Dumoulin, A. Courville, FiLM: Visual reasoning with a general conditioning layer. Proc. AAAI Conf. Artif. Intell. **32**(1) (2018). https://doi.org/10.1609/aaai.v32i1.11671
7. E. Cano, D. FitzGerald, A. Liutkus, M.D. Plumbley, F.R. Stöter, Musical source separation: An introduction. IEEE Signal Process. Mag. **36**(1), 31–40 (2019). https://doi.org/10.1109/MSP.2018.2874719
8. N.Q.K. Duong, E. Vincent, R. Gribonval, Under-determined reverberant audio source separation using a full-rank spatial covariance model. IEEE Trans. Audio Speech Lang. Process. **18**(7), 1830–1840 (2010)
9. D. FitzGerald, A. Liutkus, R. Badeau, in *Proc. of IEEE International Conference on Acoustics, Speech and Signal Processing (ICASSP)*. PROJET — Spatial audio separation using projections (Institute of Electrical and Electronics Engineers (IEEE), Shanghai, 2016), pp. 36–40
10. A. Liutkus, D. Fitzgerald, R. Badeau, in *Proc. of IEEE Workshop on Applications of Signal Processing to Audio and Acoustics (WASPAA)*. Cauchy nonnegative matrix factorization (Institute of Electrical and Electronics Engineers (IEEE), New Paltz, 2015), pp. 1–5
11. J. Le Roux, J.R. Hershey, F. Weninger, in *Proc. of IEEE International Conference on Acoustics, Speech and Signal Processing (ICASSP)*. Deep NMF for speech separation (Institute of Electrical and Electronics Engineers (IEEE), South Brisbane, 2015), pp. 66–70
12. Y. Mitsufuji, S. Koyama, H. Saruwatari, in *Proc. of IEEE International Conference on Acoustics, Speech and Signal Processing (ICASSP)*. Multichannel blind source separation based on non-negative tensor factorization in wavenumber domain (Institute of Electrical and Electronics Engineers (IEEE), Shanghai, 2016), pp. 56–60
13. A. Liutkus, D. Fitzgerald, Z. Rafii, B. Pardo, L. Daudet, Kernel additive models for source separation. IEEE Trans. Signal Process. **62**(16), 4298–4310 (2014)
14. A. Ozerov, C. Fevotte, Multichannel nonnegative matrix factorization in convolutive mixtures for audio source separation. IEEE Trans. Audio Speech Lang. Process. **18**(3), 550–563 (2010)
15. A. Liutkus, D. Fitzgerald, Z. Rafii, in *Proc. of IEEE International Conference on Acoustics, Speech and Signal Processing (ICASSP)*. Scalable audio separation with light kernel additive modelling (Institute of Electrical and Electronics Engineers (IEEE), South Brisbane, 2015), pp. 76–80
16. C. Van Der Malsburg, in *Brain Theory*, ed. by G. Palm, A. Aertsen. Frank Rosenblatt: Principles of Neurodynamics: Perceptrons and the Theory of Brain Mechanisms, (Springer Berlin Heidelberg, Berlin, Heidelberg, 1986), pp. 245–248
17. K. Fukushima, Neocognitron: A self-organizing neural network model for a mechanism of pattern recognition unaffected by shift in position. Biol. Cybern. **36**, 193–202 (1980)
18. D.E. Rumelhart, G.E. Hinton, R.J. Williams, Learning representations by back-propagating errors. Nature **323**(6088), 533–536 (1986)
19. A.A. Nugraha, A. Liutkus, E. Vincent, in *Proc. of 24th European Signal Processing Conference (EUSIPCO)*. Multichannel music separation with deep neural networks (Institute of Electrical and Electronics Engineers (IEEE), Budapest, 2016), pp. 1748–1752
20. S. Uhlich, F. Giron, Y. Mitsufuji, in *Proc. of IEEE International Conference on Acoustics, Speech and Signal Processing (ICASSP)*. Deep neural network based instrument extraction from music (Institute of Electrical and Electronics Engineers (IEEE), South Brisbane, 2015), pp. 2135–2139
21. N. Takahashi, Y. Mitsufuji, in *Proc. of IEEE Workshop on Applications of Signal Processing to Audio and Acoustics (WASPAA)*. Multi-scale multi-band densenets for audio source separation (Institute of Electrical and Electronics Engineers (IEEE), New Paltz, 2017), pp. 21–25
22. S. Uhlich, M. Porcu, F. Giron, M. Enenkl, T. Kemp, N. Takahashi, Y. Mitsufuji, in *Proc. of IEEE International Conference on Acoustics, Speech and Signal Processing (ICASSP)*. Improving music source separation based on deep neural networks through data augmentation and network blending (Institute of Electrical and Electronics Engineers (IEEE), New Orleans, 2017), pp. 261–265
23. N. Takahashi, N. Goswami, Y. Mitsufuji, in *Proc. of IWAENC*. MMDenseLSTM: An efficient combination of convolutional and recurrent neural networks for audio source separation (2018)
24. F.R. Stöter, S. Uhlich, A. Liutkus, Y. Mitsufuji, Open-Unmix - A reference implementation for music source separation. J. Open Source Softw. **4**, 1667 (2019)
25. J.H. Kim, J. Yoo, S. Chun, A. Kim, J.W. Ha, Multi-domain processing via hybrid denoising networks for speech enhancement. arXiv (2018)
26. J. Su, Z. Jin, A. Finkelstein, in *Proc. of Interspeech*. HiFi-GAN: High-Fidelity Denoising and Dereverberation Based on Speech Deep Features in Adversarial Networks (International Speech Communication Association (ISCA), Shanghai, 2020), pp. 4506–4510. https://doi.org/10.21437/Interspeech.2020-2143
27. N. Wiener, *Extrapolation, Interpolation, and Smoothing of Stationary Time Series: With Engineering Applications* (The MIT Press, USA, 1949). https://doi.org/10.7551/mitpress/2946.001.0001
28. J. Lee, Y. Jung, M. Jung, H. Kim, in *Proc. of Asia-Pacific Signal and Information Processing Association Annual Summit and Conference (APSIPA ASC)*. Dynamic noise embedding: Noise aware training and adaptation for speech enhancement (Institute of Electrical and Electronics Engineers (IEEE), Auckland, 2020), pp. 739–746
29. H. Fang, G. Carbajal, S. Wermter, T. Gerkmann, in *Proc. of IEEE International Conference on Acoustics, Speech and Signal Processing (ICASSP)*. Variational autoencoder for speech enhancement with a noise-aware encoder (Institute of Electrical and Electronics Engineers (IEEE), Toronto, 2021), pp. 676–680
30. R. Sawata, S. Uhlich, S. Takahashi, Y. Mitsufuji, in *Proc. of IEEE International Conference on Acoustics, Speech and Signal Processing (ICASSP)*. All for one and one for all: Improving music separation by bridging networks (Institute of Electrical and Electronics Engineers (IEEE), Toronto, 2021), pp. 51–55
31. A. Défossez, in *Proc. of the International Society for Music Information Retrieval (ISMIR) Conference Workshop on Music Source Separation*. Hybrid Spectrogram and Waveform Source Separation (International Society for Music Information Retrieval (ISMIR), 2021)
32. S. Rouard, F. Massa, A. Défossez, in *Proc. of IEEE International Conference on Acoustics, Speech and Signal Processing (ICASSP)*. Hybrid transformers for music source separation (IEEE, Rhodes Island, 2023), pp. 1–5
33. Y. Luo, J. Yu, Music source separation with band-split rnn. IEEE/ACM Trans. Audio Speech Lang. Process. (2023)
34. N. Takahashi, Y. Mitsufuji, in *Proc. of IEEE/CVF Conference on Computer Vision and Pattern Recognition (CVPR)*. Densely connected multidilated convolutional networks for dense prediction tasks (2021), pp. 993–1002. https://doi.org/10.1109/CVPR46437.2021.00105
35. N. Takahashi, Y. Mitsufuji, D3Net: Densely connected multidilated DenseNet for music source separation. CoRR. abs/2010.01733 (2020). https://arxiv.org/abs/2010.01733
36. F.R. Stöter, A. Liutkus, N. Ito, in *Proc. of International Conference on Latent Variable Analysis and Signal Separation*. The 2018 signal separation evaluation campaign (Springer International Publishing, Guildford, 2018), pp. 293–305
37. F. Lluís, J. Pons, X. Serra, in *Proc. of Interspeech*. End-to-end music source separation: Is it possible in the waveform domain? (International Speech Communication Association (ISCA), Graz, 2019), pp. 4619–4623
38. D. Stoller, S. Ewert, S. Dixon, in *Proc. of the 19th International Society for Music Information Retrieval (ISMIR) Conference*. Wave-U-Net: A multi-scale neural network for end-to-end audio source separation (International Society for Music Information Retrieval (ISMIR), Paris, 2018), pp. 334–340
39. A. Défossez, N. Usunier, L. Bottou, F. Bach, Demucs: Deep Extractor for Music Sources with extra unlabeled data remixed. CoRR. abs/1909.01174 (2019). http://arxiv.org/abs/1909.01174
40. Y.N. Dauphin, A. Fan, M. Auli, D. Grangier, in *Proc. of the 34th International Conference on Machine Learning (ICML)*, vol. 70. Language modeling with gated convolutional networks (Proceedings of Machine Learning Research (PMLR), Sydney, 2017), pp. 933–941
41. M. Kim, W. Choi, J. Chung, D. Lee, S. Jung, in *Proc. of the International Society for Music Information Retrieval (ISMIR) Conference Workshop on Music Source Separation*. KUIELab-MDX-Net: A two-stream neural network for music demixing (International Society for Music Information Retrieval (ISMIR), 2021)
42. C.Y. Yu, K.W. Cheuk, in *Proc. of the International Society for Music Information Retrieval (ISMIR) Conference Workshop on Music Source Separation*. Danna-Sep: Unite to separate them all (International Society for Music Information Retrieval (ISMIR), 2021)
43. W. Choi, M. Kim, J. Chung, S. Jung, in *Proc. of IEEE International Conference on Acoustics, Speech and Signal Processing (ICASSP)*. LaSAFT: Latent source attentive frequency transformation for conditioned source separation (Institute of Electrical and Electronics Engineers (IEEE), Toronto, 2021), pp. 171–175
44. Y.S. Jeong, J. Kim, W. Choi, J. Chung, S. Jung, in *Proc. of the International Society for Music Information Retrieval (ISMIR) Conference Workshop on Music Source Separation*. LightSAFT: Lightweight latent source aware frequency transform for source separation (International Society for Music Information Retrieval (ISMIR), 2021)
45. H. Liu, Q. Kong, J. Liu, in *Proc. of the International Society for Music Information Retrieval (ISMIR) Conference Workshop on Music Source Separation*. CWS-PResUNet: Music source separation with channel-wise subband phase-aware ResUNet (International Society for Music Information Retrieval (ISMIR), 2021)
46. W. Choi, M. Kim, J. Chung, D. Lee, S. Jung, in *Proc. of the 21st International Society for Music Information Retrieval (ISMIR) Conference*. Investigating U-Nets with various intermediate blocks for spectrogram-based singing voice separation (International Society for Music Information Retrieval (ISMIR), Montreal, 2020), pp. 192–198
47. O. Ronneberger, P. Fischer, T. Brox, in *Proc. of Medical Image Computing and Computer-Assisted Intervention (MICCAI)*. U-Net: Convolutional networks for biomedical image segmentation (Springer, Munich, 2015), pp. 234–241
48. Q. Kong, Y. Cao, H. Liu, K. Choi, Y. Wang, in *Proc. of the 22nd International Society for Music Information Retrieval (ISMIR) Conference*. Decoupling magnitude and phase estimation with deep ResUNet for music source separation (International Society for Music Information Retrieval (ISMIR), 2021), pp. 342–349
49. N. Takahashi, P. Agrawal, N. Goswami, Y. Mitsufuji, in *Proc. Interspeech*. PhaseNet: Discretized phase modeling with deep neural networks for audio source separation (International Speech Communication Association (ISCA), Hyderabad, 2018), pp. 2713–2717
50. D. Yin, C. Luo, Z. Xiong, W. Zeng, in *Proc. AAAI*. Phasen: A phase-and-harmonics-aware speech enhancement network (AAAI Press, New York City, 2020), pp. 9458–9465
51. T. Peer, S. Welker, T. in *Proc. of IEEE International Conference on Acoustics, Speech and Signal Processing (ICASSP)*. Gerkmann, DiffPhase: Generative Diffusion-Based STFT Phase Retrieval (Institute of Electrical and Electronics Engineers (IEEE), Rhodes Island, 2023), pp. 1–5. https://doi.org/10.1109/ICASSP49357.2023.10095396
52. H.S. Choi, J.H. Kim, J. Huh, A. Kim, J.W. Ha, K. Lee, Phase-aware Speech Enhancement with Deep Complex U-Net. CoRR abs/1903.03107 (2019). http://arxiv.org/abs/1903.03107
53. A. Vaswani, N. Shazeer, N. Parmar, J. Uszkoreit, L. Jones, A.N. Gomez, L.U. Kaiser, I. Polosukhin, in *Proc. of Advances in Neural Information Processing Systems (NeurIPS)*, ed. by I. Guyon, U.V. Luxburg, S. Bengio, H. Wallach, R. Fergus, S. Vishwanathan, R. Garnett. Attention is all you need, vol. 30 (Neural Information Processing Systems Foundation, Inc. (NeurIPS), Long Beach, 2017)
54. J. Hu, L. Shen, G. Sun, in *Proc. of the IEEE Conference on Computer Vision and Pattern Recognition (CVPR)*. Squeeze-and-excitation networks (Institute of Electrical and Electronics Engineers (IEEE), Salt Lake City, 2018)
55. Y. Luo, Z. Chen, N. Mesgarani, T. Yoshioka, in *Proc. of IEEE International Conference on Acoustics, Speech and Signal Processing (ICASSP)*. End-to-end microphone permutation and number invariant multi-channel speech separation (Institute of Electrical and Electronics Engineers (IEEE), Barcelona, 2020), pp. 6394–6398
56. R. Caruana, Multitask learning. Mach. Learn. **28**(1), 41–75 (1997)
57. Z. Rafii, A. Liutkus, F.R. Stöter, S.I. Mimilakis, R. Bittner. The MUSDB18 corpus for music separation (2017). https://doi.org/10.5281/zenodo.1117372
58. D.P. Kingma, J. Ba, in *Proc. of 3rd International Conference on Learning Representations (ICLR)*, ed. by Y. Bengio, Y. LeCun. Adam: A method for stochastic optimization (OpenReview.net, San Diego, 2015)

**Publisher's Note**